\documentclass[aps,showpacs,nofootinbib,preprint]{revtex4}
\usepackage{epsf,amsmath,amssymb,subfigure}
\usepackage{epsfig}
\usepackage{graphicx,latexsym}
\usepackage{psfrag}
\usepackage[english]{babel}

\newcommand{\br}{\mathbf{r}}

\newcommand{\bq}{\mathbf{q}}

\newcommand{\hq}{\tilde h_\bq}

\newcommand{\SC}{_{SC}}
\newcommand{\lj}{_{LJ}}

\newcommand{\rij}{r_{ij}}

\newcommand{\eij}{\epsilon_{\alpha \beta}}

\newcommand{\chd}{\text{CH}_2}
\newcommand{\cht}{\text{CH}_3}

\newcommand{\qm}{{q_{min}}}

\newcommand{\qs}{{q^*}}

\newcommand{\media}[1]{\left\langle  #1 \right\rangle}
\newcommand{\p}[1]{\left({#1}\right)}
\newcommand{\pq}[1]{\left[{#1}\right]}

\newcommand{\um}{\frac{1}{2}}
\newcommand{\de}{{\mathrm d}}
\newcommand{\D}[2]{\frac{\partial #1}{\partial #2}}

\newcommand{\dep}[2]{\frac{\partial #1}{\partial #2}}

\begin{document}
\title{Surface tension in bilayer membranes with fixed projected area}
\author{Alberto Imparato}
\affiliation{ Dipartimento di Scienze Fisiche and INFN-Sezione di Napoli,
Universit\`a ``Federico II'', Complesso Universitario di Monte S.
Angelo, I-80126 Napoli (Italy).}
\email{imparato@na.infn.it}
\begin{abstract}
We study  the elastic response of bilayer membranes with fixed projected area to both stretching and
shape deformations. A surface tension is associated to each of these deformations.
By using model amphiphilic membranes and computer simulations, we are able to 
observe both the types of deformation, and thus,
both the surface tensions, related to each type of deformation, are measured for the same system.
These surface tensions are found to assume different values in the same bilayer membrane: in particular they vanish for different values of the projected area.
We introduce a simple theory which relates the two quantities and successfully apply it to the data obtained with computer simulations.
\end{abstract}
\maketitle
 
\maketitle
Bilayer membranes are composed of amphiphilic molecules whose hydrophobic part is strongly insoluble in water. Such molecules tends to form interfaces with the solvent in order to reduce the interfacial energy \cite{Isr}: a bilayer
is thus formed by two adjacent sheets of amphiphilic molecules separating two aqueous phases. The interest in such structures resides in the fact that they 
play an outstanding role in the organization of biological cells:
bilayer membranes of lipids form the basic structure
of a cell's plasma membrane and of internal membranes, which
surround the organelles in eukaryotic cells, such as the nucleus,
 mitochondria and the Golgi apparatus \cite{ABL,L1}.

In a fluid membrane fluctuating freely in a solvent, all                        internal degrees of freedom, related, e.g, to the hydrocarbon chain conformation,
and to to the local molecular density,  relax on a fast time scale.                  The membrane is then characterized by a zero shear modulus, i.e.
any shear deformation induces a flow of the amphiphilic molecules within the membrane.
Such a system is then subject to only two types of elastic deformations:
bending deformations and stretching deformations.
The first are deformations
normal to the membrane plane which change the membrane shapes,  while the latter are in-plane deformations which change the local
area per molecule projected onto the membrane midplane.

The surface of a fluctuating membrane can thus be viewed as an interface with the solvent, and we will call {\it fluctuation} surface tension the surface tension of the bilayer membrane associated to shape fluctuations.
The surface tension of an interface saturated by surfactant molecules is expected
to vanish (see, e.g., \cite{Saf}), and thus the fluctuation surface tension of a fluid membrane fluctuating freely in a solvent is usually assumed to be zero.
However in the case of a constrained bilayer membrane, such as those fixed on a frame, the geometrical constrain can lead to a non vanishing surface tension.
Non vanishing fluctuation surface tension have been observed, e.g., in \cite{MM}.

On a macroscopic length scale, a membrane fixed on a frame can
be treated as an elastic sheet: if one of the sides of the frame
is movable and one tries to change the frame area by
pulling or pushing on this side, an elastic force will appear
which will tend to restore the preferred value of the frame area. 
The parameter describing such elastic behaviour is the {\it mechanical} surface tension
and can assume non negative values too \cite{GL1,LE2}.

Intuitively, one would expect that the two surface tensions associated with the two different membrane deformations, namely shape deformation and lateral stretching, are different quantities, see ref.~\cite{DL} and discussion therein. However, in experiments, they are usually measured by observing
the one or the other  type of deformation, and so one cannot compare them experimentally for the same system. On the contrary, the two  surface tensions can be measured using computer simulations of model membranes, and this has
been done in two previous works \cite{FP0,Ste}. But in one case, the two surface tensions have been found to be equal within the statistical errors \cite{FP0}, while in the other case they have been found to be proportional with respect to each other \cite{Ste}.

The aim of this paper is thus the systematical study and comparison between the fluctuation  surface tension and the mechanical surface tension, measured by observing both the types 
of deformation in the same bilayer membrane.
In the present paper we consider bilayer membrane with fixed projected area. The projected area correspond to the area of the frame in the picture drawn above.
 By using model bilayer membrane and computer simulations, we measure both the surface tension associated to shape deformation and that associated to stretching deformations. 
We show, for the first time,  that in the same system these two quantities assume different values, and in particular they do not vanish for the same 
value of the projected area.
The model used here for the computer simulations, have been largely used in previous works and
has been quite  successful in describing many dynamic and elastic properties of 
real bilayer membranes \cite{GL1,GL2,alb1,alb2}.

The paper is organized as follows: in section \ref{model}, we describe the model
bilayer, together with the computer simulation techniques which we use in the present work.
In section \ref{mis_S} we discuss the theory of the mechanical surface tension 
and the method adopted to measure it. We then show the results for such quantity obtained by computer simulations.
In section \ref{mis_s}, after reviewing the classical elasticity theory for the shape fluctuations of bilayer membranes, we show the results for the shape fluctuation surface tension, obtained by computer simulations.
In section \ref{compare_sec}, the two quantities are compared, and a simple theory which connects them is introduced and discussed. 
We discuss our results and conclude in section \ref{concl}.

In the following, we consider a bilayer membrane made up of $N$ amphiphilic molecules, at temperature $T$, spanning a square frame of area $A_p=L^2$, its effective area being $A$.
\section{Model bilayers and computer simulations}\label{model}
\subsection{Coarse grained model of amphiphilic bilayers}
\begin{figure}[h]
\center
\psfrag{fi}[.9]{$\phi_i$}
\psfrag{i1}[.9]{$i-1$}
\psfrag{i2}[.9]{$i$}
\psfrag{i3}[.9]{$i+1$}
\includegraphics[width=3cm]{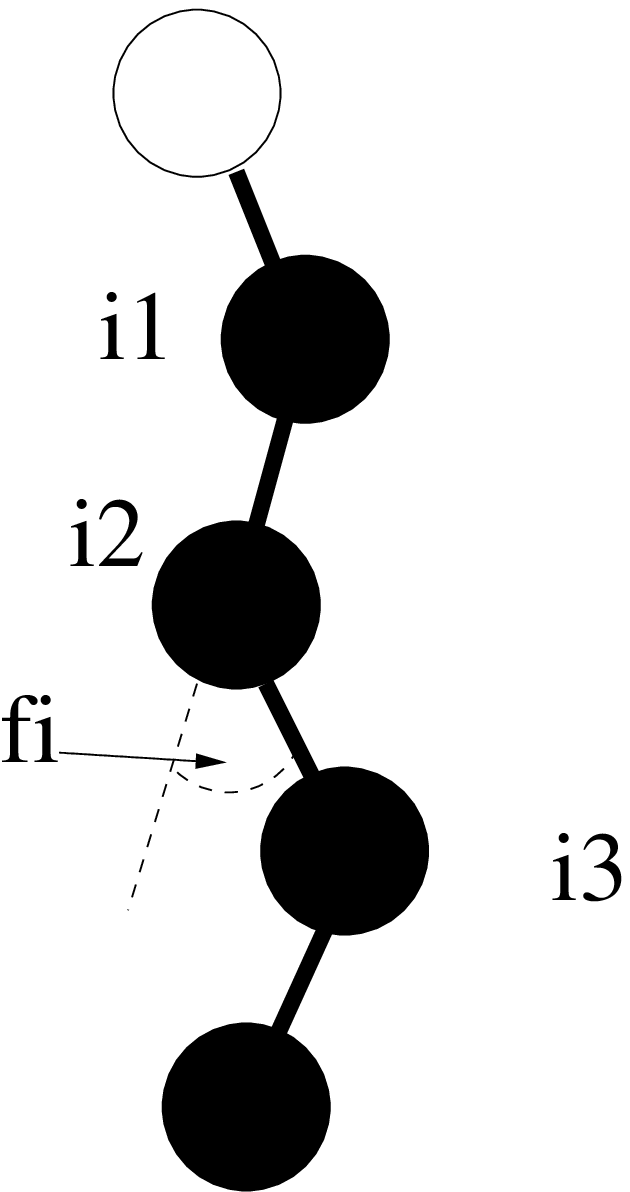}
\caption{Cartoon of the model amphiphile molecule used in this paper.}
\label{mol}
\end{figure}

We adopt the same coarse grained model introduced and used in refs.\cite{GL1,GL2,alb1,alb2}, 
which turned out to be an effective model to study several properties of bilayer membranes such as surface tension, bending rigidity and diffusion characteristics.
The amphiphilic molecules are represented as linear chains of beads, a single bead representing the molecule hydrophilic head (H), or several $\mathrm{CH}_2/ \mathrm{CH}_3$ groups of the amphiphile hydrocarbon chain (C) see fig.~\ref{mol}.  The water molecule is also represented as a single bead (W).
The hydrophobic interaction of C with W and H particles is modelled by soft core potential
\begin{equation}
U\SC(\rij)=4 \eij \p {\frac{ \ell'}{\rij}}^9 \, ,
\label{usc}
\end{equation}
while attractive interactions W-W, W-H, H-H, C-C, are modelled by a Lennard-Jones potential
\begin{equation}
U\lj(\rij)=4 \eij \pq { \p{\frac{\ell}{\rij}}^{12}-\p{\frac{\ell}{\rij}}^{6}} \, ,
\label{ulj}
\end{equation}
where $\alpha/\beta\, \epsilon \, \{ \text W, \text H, \text C \} $.
Adjacent beads along a single model molecule interact via the harmonic potential
\begin{equation}
U_2(r_{i,i+1})=k_2\p{r_{i,i+1}-\ell}^2\, ,
\label{u2}
\end{equation}
where $i$ and $i+1$ indicate two successive particles along the chain.
The effects of hydrocarbon chain stiffness is modelled by letting all the particles within a single amphiphile molecule interact  via the three-body bending potential
\begin{equation}
U_3(\br_{i-1,i},\br_{i,i+1})=k_3\p{1-\frac{\br_{i-1,i}\cdot \br_{i,i+1}}{r_{i-1,i}\, r_{i,i+1}}}=k_3(1-\cos \phi_i) \, ,
\label{u3}
\end{equation}
where 
$\br_{i,i+1}=\br_{i+1}-\br_i$, see fig. \ref{mol}.

\subsection{Simulation method and model parameters}
In the present work we combine both Monte-Carlo (MC) and Molecular Dynamics (MD) simulation methods.
Our simulations are carried out for a cuboidal boxes of constant volume
and periodic boundary conditions.
The MC algorithm is used to make the system relax towards configurations of minimal potential energy, the output of $500\times N$ MC steps are used as starting configurations for
the MD part of the simulations.

For the parameters characterizing the above potentials we choose the values 
used in refs.~\cite{GL1,GL2,alb1,alb2}. For the interaction ranges of the LJ potentials we take $\ell=1/3$ nm 
 which is of the same order as
the LJ lengths of interactions for pairs of $\chd$ groups, $\cht$
groups or water molecules, as discussed  in \cite{EB}.
The characteristic length of the repulsive SC potential (\ref{u3}), is taken to be  $\ell'=1.05 \ell$ as in \cite{GL1}: with this choice the hard-core repulsion of the SC potential is of the same order of the repulsive part of the LJ potential. 
We take  $\epsilon=2/N_{A_{V}}$ kJ ($N_{A_{V}}$ is the Avogadro number): this value is bigger than the LJ energies for pairs of $\chd$ and/or water
molecules reported in \cite{EB}, but it takes into account that in our model
one C particle corresponds to three or four $\chd$ groups \cite{GL1}.
The strength of the harmonic bond potential (\ref{u2}) and of the three-body bending potentials (\ref{u3}),
 are taken to be $k_2=5000 \epsilon/\ell^2$ and $k_3=2 \epsilon$  \cite{GL1}.

Since we run MD simulations, two additional parameters are involved in the present model: the masses of the different beads, which enter the equations of motions, and the 
time step $\delta t$, used to discretize such equations.
For simplicity's sake, all the beads are taken to have  the same mass
 $m=0.036/N_{A_{V}}$ kg,
which is approximately the mean value of the mass of a water molecule and  the mass of four $\chd$ groups.
Using this value of $m$ one obtains the characteristic time scale $\tau=\sqrt{m \ell^2 /\epsilon}=1.4$ ps.
In the MD simulations, the integration time step $\delta t$ is taken to be $\delta t=\tau/2000=0.7$ fs.
The {\it leap-frog} algorithm at constant temperature \cite{AT} with $k_B T=1.35 \epsilon$ (corresponding to $T=325$~K) is used for the integration of the equations of motions.
In the following, if not differently specified, all the quantities will be expressed in units of the tree basic parameters $\ell$, $\epsilon$ and $m$.

In the present paper we consider bilayers with three different values of the number of amphiphilic molecules $N=512$, $N=768$  and $N=1152$.
For each value of $N$, the simulations are run with different values of the projected area $A_p$, which corresponds to the area of the simulation box side parallel to the bilayer.
In changing the projected area, we keep constant the number of water particles $N_W$, as well as the overall simulation box volume $V$, i.e., we keep the overall system density constant.
The values of $N$, $N_W$ and $V$ used in the simulations described in the following are reported in table \ref{tab1}.
\begin{table}[h]
\center
\begin{tabular} {|c|c|c|}
\hline
 $N$&  $N_W$&  $V(\ell^3)$ \\
512 & 3200 & 8640\\
768 & 4800&   12960\\
1152 & 7200 & 19440\\ 
\hline
\end{tabular}
\caption{Number of amphiphiles $N$, number of water particles $N_W$ and  simulation box volume $V$ in $\ell^3$ units, used in the simulations described in the text.}
\label{tab1}
\end{table}
\section{Stress tensor and mechanical surface tension}\label{mis_S}
On a macroscopic length scale, a membrane fixed on a frame can be viewed as an elastic sheet: the system will be characterized by some equilibrium value of the frame area where the force exerted on the frame sizes vanishes, and small changes in  
the frame area cause restoring forces to appear.
If $F$ is the free energy of a framed bilayer, the surface tension $\Sigma$ conjugated
to its projected area $A_p$ is defined as 
\begin{equation}
\Sigma=\dep{F}{A_p}.
\label{defSigma}
\end{equation} 

In the present work, the bilayer surface tension is measured using a method
first introduced
by Schofield and Henderson \cite{SH},
and then extended by Goetz and Lipowsky \cite{GL1}.
A fluctuating membranes can be considered isotropic and homogeneous along any direction parallel to the membrane plane, if the amplitude of the fluctuations is not too high.
Thus, the system stress tensor will be a function of the coordinate $z$ perpendicular to the bilayer plane. 
The surface tension is related to the system stress tensor via
\begin{equation}
\Sigma=\int \de z \pq{\Sigma_T(z) -\Sigma_N(z)},
\label{defS}
\end{equation} 
where $\Sigma_T(z)$ and $\Sigma_N(z)$ are the components of the stress tensor perpendicular and normal to the bilayer surface, respectively.
The integral can be extended to infinity because the stress tensor in isotropic in the  bulk water and so $\Sigma_T(z)=\Sigma_N(z)$.
By noting that the stress tensor is defined as the negative of the pressure tensor, one sees that eq.~(\ref{defS}) is equivalent to the 
 original expression of the surface tension of a planar
liquid-vapor interface, as formulated by Kirkwood and Buff \cite{KB}.
The macroscopic stress tensor  $\mathbf \Sigma$ can be expressed in terms of 
the microscopic stress $\mathbf s$ which depends on the momenta and on the positions of the system particles within a small volume. 
The microscopic and the macroscopic stress tensor are related via
\begin{equation}
{\mathbf \Sigma}=\media{\mathbf s},
\label{medias}
\end{equation} 
where the brackets denote ensemble average.
The microscopic stress tensor ${\mathbf s}$ consists of two contributions
one arising from the   kinetic energy and the other from the interaction energy of the system particles within a small volume ${\mathbf s}={\mathbf s}_K+{\mathbf s}_{in}$.
The kinetic contribution ${\mathbf s}_K$ to the macroscopic stress tensor, 
as given by eq.~(\ref{medias}), can be neglected, since it is isotropic, and must vanish on average. 
Thus, one is left with the interaction part of the microscopic stress tensor ${\mathbf s}_{in}$ only, and eq. (\ref{medias}) becomes
\begin{equation}
{\mathbf \Sigma}=\media{\mathbf s}_{in},
\label{mediasin}
\end{equation}
Furthermore a planar bilayer membrane in solvent can be considered  homogeneous along any direction
parallel to the the bilayer plane, and so is the stress tensor. The system 
can thus be divided in thin slices parallel to the bilayer plane, and the stress tensor can be averaged over each of these slices.
Thus the microscopic stress tensor is given by  \cite{GL1,SH,LE2}
\begin{equation}
{\mathbf s}_{in}(z)=f(z_i,z_j,z) \frac{1}{\Delta V}\sum_{i>j} {\mathbf F}_{ij}\otimes {\mathbf r}_{ij}, 
\label{defsz}
\end{equation} 
where the sum is extended to all the system particle pairs,  $\Delta V$ is the volume of the slice,  the function $f(z_i,z_j,z)$ determines the actual contribution of the pair $i,j$ to the stress tensor in the slice of coordinate $z$, and the symbol $\otimes$ denotes the tensorial product.
If both the particles are inside the current slice we take $f(z_i,z_j,z)=1$ \footnote[2]{In ref.\cite{GL1}, the expression for the contribution to the stress tensor of the three body interaction (\ref{u3}) was sligthly different. However, a three body interaction can be expressed as the sum of pairwise interactions \cite{AT}, and thus eq.(\ref{defsz}) is correct also for the three body potential (\ref{u3}).}.
Let $\Delta z$ be the thickness of each slice, 
if both the particles are external to the current slice and on opposite sides, 
we take  $f(z_i,z_j,z)=\Delta z/|z_i-z_j|$, where, using periodic boundary conditions, the shortest distance $|z_i-z_j|$ between them is considered. If just one of the two particles lies within the current slice, we take  $f(z_i,z_j,z)=\Delta z/(2|z_i-z_j|)$. In all the other cases, the contribution of the particles $i,j$ to the 
 stress tensor  associated to the slice of coordinate $z$ vanishes, and thus we take $f(z_i,z_j,z)=0$.
Inspection of eq.~(\ref{defsz}) suggests that its rhs can be viewed as a local density  of  the macroscopic virial
tensor ${\mathbf W}=\sum_{i>j} {\mathbf F}_{ij}\otimes {\mathbf r}_{ij}$, and thus the function $f(z_i,z_j,z)$
determines the fraction of virial tensor density to be associated to the slice 
 of coordinate $z$.
\subsection{Measurements of the microscopic stress tensor}
Using the model amphiphilic bilayer described in section \ref{model}, the microscopic stress tensor $\mathbf s_{in}(z)$, as given by eq.~(\ref{defsz}), was measured for the three values of the number of amphiphilic molecules $N$ here considered, $N=512$, $N=768$, and $N=1152$. For each value of 
$N$, $\mathbf s_{in}(z)$ was measured for different values of the bilayer projected area $A_p$ keeping constant the total volume $V$, see table  \ref{tab1}.
The simulation boxes have been divided into 140 $z$-slices, which corresponds to $\Delta z\sim 0.1\ell$, for the simulation box dimensions here used.
The off-diagonal elements of the system stress tensor are found to vanish as expected, as well as the average of its kinetic part (data not shown). 
For each value of $A_p$,  the mechanical surface tension $\Sigma$ is obtained  using eq.~(\ref{defS}), by performing a discrete integration.  
For each value of $N$ and $A_p$ we run 5 simulations of $2\times 10^5$ MD steps, and a run of $6\times 10^6$ MC steps was interposed after each run of $2\times 10^5$ MD steps.  The MC steps are inserted to allow the system to reach regions of its phase space not easily accessible with a single MD trajectory.

It is worth noting that the code for the stress tensor measuring is extremely time-consuming: a single run of $2\times 10^5$ MD steps, for a system with $N=1152$ amphiphiles, needs about one month of machine time, on a computer equipped 
with a 1 GHz processor.
The relative fluctuations of the
surface tension with respect to its average value are very large, as found in other work \cite{GL1}.
Following ref. \cite{GL1}, in order to reduce the effect of short time
fluctuations, the total simulation time of each run is divided in blocks
of 5000 MD steps, and the surface tension is subaveraged over each of these
blocks. This average value over the 5000 MD steps is then used as the sample
value for the current block.

The surface tension obtained from simulations is plotted in fig.~\ref{compare_S}, as a 
function of the projected area per amphiphile $a_p=A_p/(N/2)$, for the three values of $N$ here considered.

According to the classical elasticity theory, the parameter describing the effective compressibility of an elastic sheet to a change in its projected area is the 
area compressibility, $K_A$ defined as
\begin{equation}
K_A=A_p\dep{\Sigma}{A_p} .
\end{equation} 
By integrating this last equation, one obtains
\begin{equation}
\Sigma=K_A\ln\p{\frac{A_p}{A^\dagger_p}}\simeq K_A\frac{A_p-A^\dagger_p}{A^\dagger_p},
\label{SigKA}
\end{equation} 
where the last equality holds for values of $A_p$ close to $A^\dagger_p$, which is the area at which $\Sigma$ vanishes.
Upon integration of eq.~(\ref{SigKA}), we obtain the classical expression for 
the free energy of a stretched (or compressed) elastic sheet as a a function
of $A_p$ around the equilibrium value $A^\dagger_p$
\begin{equation}
F=\frac{K_A}{2 A^\dagger_p} \p{A_p-A^\dagger_p}^2, 
\end{equation} 
where the reference free energy at $A=A^\dagger_p$ has been chosen to be equal to zero.
Inspection of figure \ref{compare_S} indicates that the surface tension $\Sigma$
obtained by simulations follows the behaviour predicted by eq.~(\ref{SigKA}):
this quantity   is linear within a range of values around the 
equilibrium area $A^\dagger_p$. 
We find that for $A_p\ll A^\dagger_p$, ($a_p\simeq2.065$) the surface tension 
$\Sigma$ is no longer a monotonous function of $a_p$: this is probably due to the fact that, for such small projected area per molecule, the system exhibits
buckling, and the membrane cannot be considered flat on average.
In this case, the basic hypothesis that the membrane is isotropic and homogeneous along the the $z$-axis is not fulfilled. However, such hypothesis is required for measuring $\Sigma$ with the method discussed in this section, 
and thus the results obtained for very small values of $a_p$ must be inaccurate. 
Thus in the following we will consider values of $a_p$ such that $\Sigma$ is a monotonic function of $a_p$, i.e. $a_p>2.065$.
The values of the equilibrium area, of the area compressibility and of the slope of $\Sigma$ as a function of $a_p$,  for the three system sizes here considered, are listed in table \ref{tabS}.
In the same table, we also report the results for a smaller system which was considered in a previous work \cite{GL1}.
\begin{table}[h]
\center
\begin{tabular} {|c|c|c|c|}
\hline
 $N$&  $a^\dagger_p\, (\ell^2)$&  $K_A\, (\epsilon/\ell^2)$& $\partial \Sigma/ \partial a_p\, (\epsilon/\ell^3)$\\
128 & $2.15\pm0.02$ & $11.8\pm1.5$& $5.5\pm 0.7$ \\
512 & $2.12\pm0.01$ & $11.7\pm1.5$& $5.5 \pm 0.7$ \\
768 & $ 2.12 \pm 0.01$ & $9.4 \pm 1.1$ & $4.5 \pm 0.5$\\
1152 & $2.12\pm0.01$& $8.8\pm0.8$& $4.1 \pm 0.4$ \\
\hline
\end{tabular}
\caption{Equilibrium projected area per amphiphile $a^\dagger_p$, area compressibility $K_A$ for different system sizes, and slope of $\Sigma$ with respect to the projected area per amphiphile $a_p$.
The values for $N=512,768,1152$ are obtained 
by linear fit of the data shown in fig.~\ref{compare_S} to eq.(\ref{SigKA}), in the range where $\Sigma$ is a linear function of $a_p$.
The data for $N=128$ are taken by ref.~\cite{GL1}.}
\label{tabS}
\end{table}
The results listed in table \ref{tabS} suggests that the equilibrium area $A^\dagger_p$ does not depend on the system size, but is an intrinsic properties of 
the model amphiphilic molecule used here. However, the area compressibility 
decreases as a function of the system size, as already found in \cite{MM}:
this decrease is due to the fact that a larger number of oscillation modes 
are introduced in the system as its projected area is increased. Thus, the bilayer 
becomes easier to compress or to stretch if its  total projected area $A_p$ increases
while the projected area per amphiphile $a_p$ is kept constant.
Note that the values of the area compressibility found for the present model range between 263 and 360 $mJ/m^2$, see table \ref{tabS}. Such values are slightly greater than those observed for real lipid bilayers, that are in the range 140-240 $mJ/m^2$ \cite{Ma}, but are very similar to those values found in other works on computer simulations of amphiphilic bilayer with atomic resolution \cite{MM,LE2}

\section{Elasticity Hamiltonian and the fluctuation surface tension.}\label{mis_s}
\begin{figure}[h]
\center
\psfrag{S}[bl][bl][1.][-90]{$S$}
\psfrag{q}[br][br][1.]{$q$}
\psfrag{34.5}[tr][tr][.8]{$a_p=2.066$}
\psfrag{34.74}[tr][tr][.8]{$a_p=2.095$}
\psfrag{35}[tr][tr][.8]{$a_p=2.127$}
\psfrag{35.25}[tr][tr][.8]{$a_p=2.157$}
\psfrag{35.625}[tr][tr][.8]{$a_p=2.203$}
\includegraphics[width=12cm]{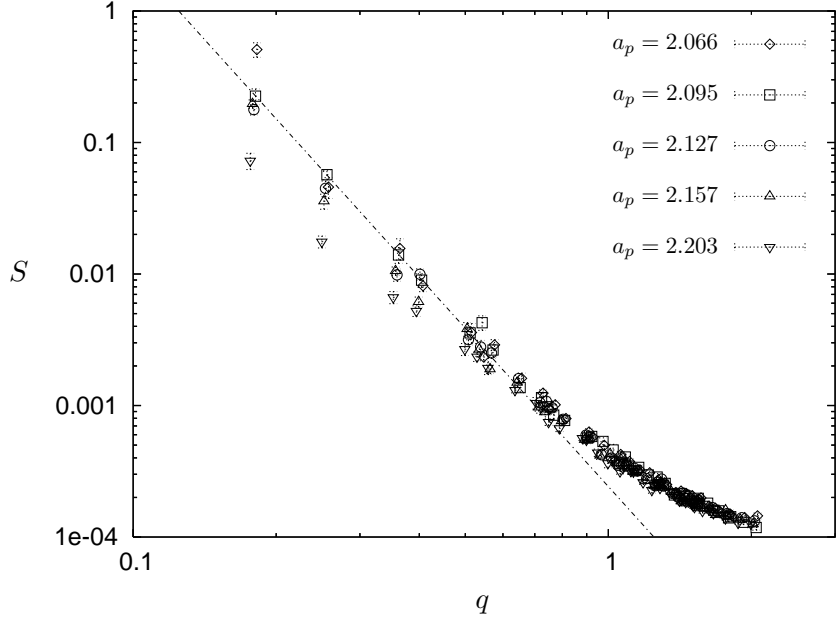}
\caption{Fluctuation spectrum $S$ as a function of the wavenumber $q$, for different values of the projected area per amphiphile $a_p$, for bilayers with $N=1152$. The tensionless state ($\sigma=0$) corresponds to the projected area per amphiphile $a_p^*=2.095$. The dashed line is obtained by fitting the fluctuation spectrum of the tensionless system  to eq. (\ref{fluspe}), with $\sigma=0$, for $q\lesssim 1$.}
\label{sq2}
\end{figure}

\begin{figure}[h]
\center
\psfrag{x3}[bl][bl][.8]{$N=1152$}
\psfrag{xm}[bl][bl][.8]{$N=768$}
\psfrag{x2}[bl][bl][.8]{$N=512$}
\psfrag{sigma}[br][br][1.][-90]{$\sigma$}
\psfrag{ap}[ct][ct][1.]{$a_p$}
\includegraphics[width=12cm]{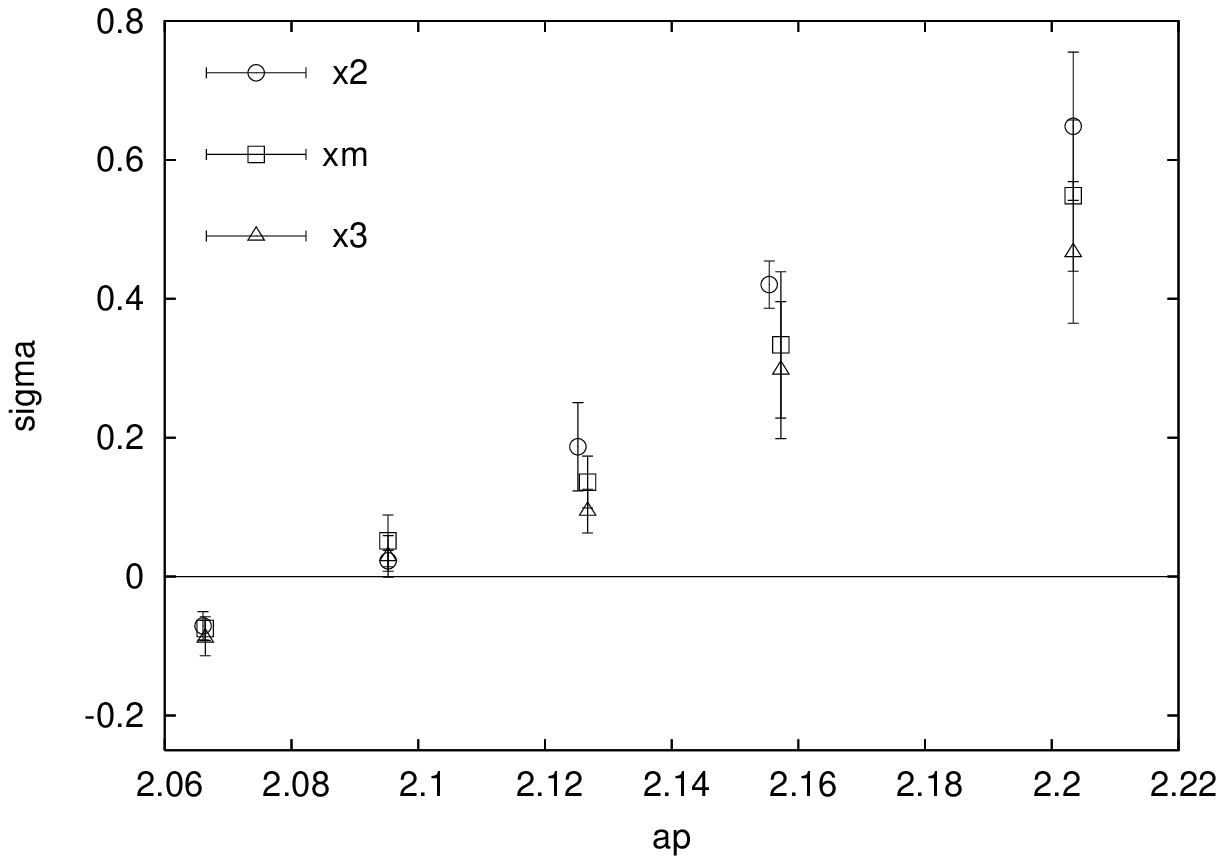}
\caption{Plot of $\sigma$ as a function of the projected area per amphiphile $a_p$ for the three values of $N$ here considered.}
\label{compare_s}
\end{figure}

In the Monge representation, the shape of the bilayer is described by the height function $h(x,y)$ which measures the distance of its midsurface from the reference $(x,y)$ plane.
The classical elasticity Hamiltonian of fluctuating membranes, for small fluctuations, reads \cite{Can,Helf}
\begin{eqnarray}
H&=&\sigma A_p +\int_{A_p} \de x \de y\pq{ \frac 1 2  \sigma (\nabla h)^2 + 
 \frac 1 2 \kappa (\nabla^2 h)^2}\nonumber \\
  &=&\sigma A_p + \frac{1} {2 L^2} \sum _{\bq}\pq{\sigma q^2+ \kappa q^4} |\hq|^2 \, ,
\label{fapp}
\end{eqnarray}
where $\hq$ define the Fourier coefficients of $h(x,y)$
\begin{equation}
\hq=\frac  {1} {A_p}\int_{A_p} \de x \de y\, \exp\pq{\mathrm{i} \br\cdot \bq} h(x,y).
\end{equation} 
The two parameters appearing in eq.~(\ref{fapp}) are the {\it bending rigidity} $\kappa$, describing the resistance of the system to bending, and the surface tension $\sigma$, which takes into account the contribution of the bilayer total area $A$ to the system energy. In fact, eq.~(\ref{fapp}) can be rewritten as
\begin{equation}
H=\sigma A+ \int \de x \de y \frac 1 2 \kappa (\nabla^2 h)^2.
\end{equation} 
The membrane fluctuation spectrum, defined as $S(q)\equiv|\hq|^2$, depends  only on $q=|\bq|$ and exhibits the functional form \cite{LB}
\begin{equation}
S(q)=\frac{k_B T}{A_p \p{\sigma q^2 +\kappa q^4}}.
\label{fluspe}
\end{equation} 
Equation (\ref{fluspe}) has been obtained under the hypothesis that the bilayer
can be modelled as a geometrical surface, neglecting its thickness.
Thus eq.~(\ref{fluspe}) holds only for wavenumbers smaller than an upper bond
$\qs$, which is usually taken to be $\qs\simeq 2 \pi /d_0$, where $d_0$ is the bilayer thickness. We find that the model bilayer thickness is $d_0\simeq 6 \ell$, independently of the system size.
For $q>q^*$, the shape of the bilayer is characterized by the local protrusions of the amphiphiles, and thus the fluctuation spectrum is dominated by a local surface tension term, see refs \cite{Ste,GL2,alb2}. 
However these sub-optical modes are not accessible in experiments, and thus, in the present work, we will focus on the small wavenumber, and on the effective values of $\kappa$ and $\sigma$ that can be obtained by eq.~(\ref{fluspe}).
By sampling the height field $h(x,y)$ during the simulations, one can obtain the mean fluctuation spectrum (\ref{fluspe}). By fitting $S(q)$ to the rhs of eq.
(\ref{fluspe}), for $q\lesssim q^*$, the values of the bending rigidity $\kappa$ and of the surface tension $\sigma$ can thus be estimated.
The fluctuation spectrum shape depends on the projected area:
and thus
both $\kappa$ and $\sigma$ depend on this parameter. 
By changing the value of $A_p$ one can achieve the tensionless ($\sigma=0$) state \cite{GL2,alb2}.
The surface tension has been measured for the three system sizes here considered $N=521, \, 768,\, 1152$, for different projected area, and keeping constant the overall volume, see table \ref{tab1}. For each value of the projected area, we run $10\times 10^5$ MD steps: a run 
of $6\times 10^6$ MC steps has been interposed after each run of $10^5$ steps.
The fluctuation spectrum has been sampled during the MD simulations, every $5000$ MD steps.
In  fig.\ref{sq2} the fluctuation spectrum of the larger system here considered is plotted as a function of the wavenumber $q$ for different value of the projected area per amphiphile $a_p$:
 the surface tension $\sigma$ vanishes for a given values 
of $a_p=a^*_p$, which depends on the microscopic details of the system (on the model parameters in the case of simulations), while for $a_p>a_p^*$ ($a_p<a_p^*$) it is greater (smaller) than zero. This is clearly shown in  figure \ref{compare_s}, where the surface tension $\sigma$ is plotted as a function of $a_p$
for the three values of $N$ here considered.

By linear fit of the data in figure \ref{compare_s}, we obtain the tensionless area $a^*_p$ and the slope of the surface tension $\sigma$ as a function of $a_p$, the results are listed in table \ref{tabsig}.
As in the case of the mechanical surface tension $\Sigma$, the slope of the curve $\sigma(a_p)$ decreases as $N$ is increased. Comparison of tables \ref{tabS} and \ref{tabsig} indicates that the slopes of $\Sigma$ and $\sigma$ as functions of $a_p$ are different for a given system size.
\begin{table}[h]
\center
\begin{tabular} {|c|c|c|}
\hline
 $N$&  $a^\dagger_p\, (\ell^2)$& $\partial \sigma/ \partial a_p\, (\epsilon/\ell^3)$\\
512 & $2.085\pm0.005$ & $5.4 \pm 0.7$ \\
768 & $ 2.09 \pm 0.01$ & $4.0 \pm 0.4$\\
1152 & $2.09\pm0.01$&  $3.6 \pm 0.4$ \\
\hline
\end{tabular}
\caption{Equilibrium projected area per amphiphile $a^*_p$, and slope of $\sigma$ with respect to the projected area per amphiphile $a_p$.
The values for the three system size are obtained 
by linear fit of the data shown in fig.~\ref{compare_s}.}
\label{tabsig}
\end{table}

\section{Comparison between the two surface tension $\sigma$ and $\Sigma$}\label{compare_sec}
\begin{figure}[h]
\center
\psfrag{x3}[bl][bl][.8]{$N=1152$}
\psfrag{xm}[bl][bl][.8]{$N=768$}
\psfrag{x2}[bl][bl][.8]{$N=512$}
\psfrag{Sigma}[br][br][1.][-90]{$\Sigma$}
\psfrag{ap}[ct][ct][1.]{$a_p$}
\includegraphics[width=12cm]{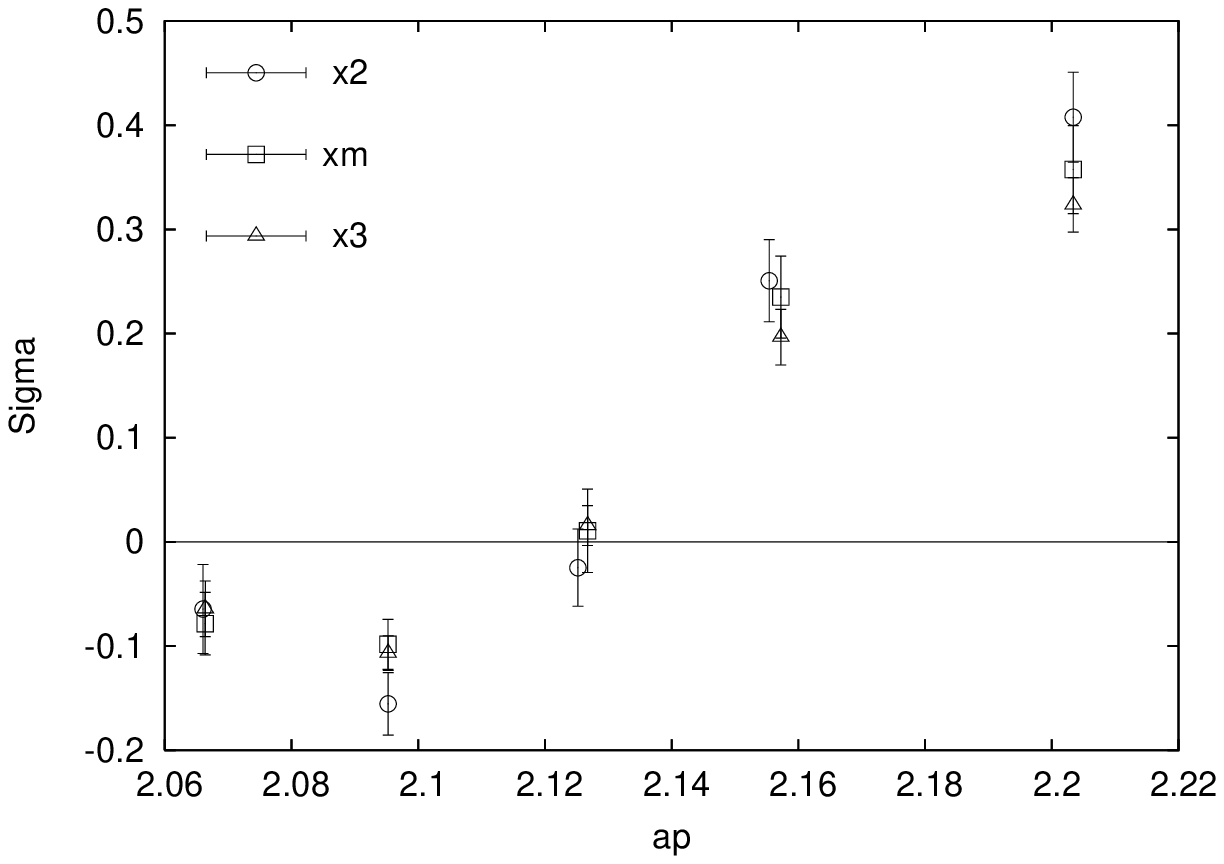}
\caption{Plot of $\Sigma$ as a function of the projected area per amphiphile $a_p$ for the three values of $N$ here considered, $N=512$, $N=768$ and $N=1152$.}
\label{compare_S}
\end{figure}

Comparison of figure \ref{compare_s} with figure \ref{compare_S} shows clearly that the projected area per amphiphile $a_p^*\simeq2.09$ where the surface tension $\sigma$ vanishes
is different from the projected area per amphiphile $a_p^\dagger\simeq2.12$, at which the mechanical tension $\Sigma$ vanishes. Since these two quantities have been measured independently, for {\it each} value of the projected area, this 
last result suggests that one is dealing with two different quantities.
In addition, the two vanishing areas  $a_p^*$ and  $a_p^\dagger$ do not exhibit any dependence on the system size $N$, and thus one can argue that they are two independent intrinsic properties of the amphiphilic molecule. 

\begin{figure}[h]
\center
\psfrag{ap}[ct][ct][1.]{$a_p$}
\psfrag{sigma}[ct][ct][.8]{$\sigma$}
\psfrag{Sigma}[ct][ct][.8]{$\Sigma$}
\includegraphics[width=12cm]{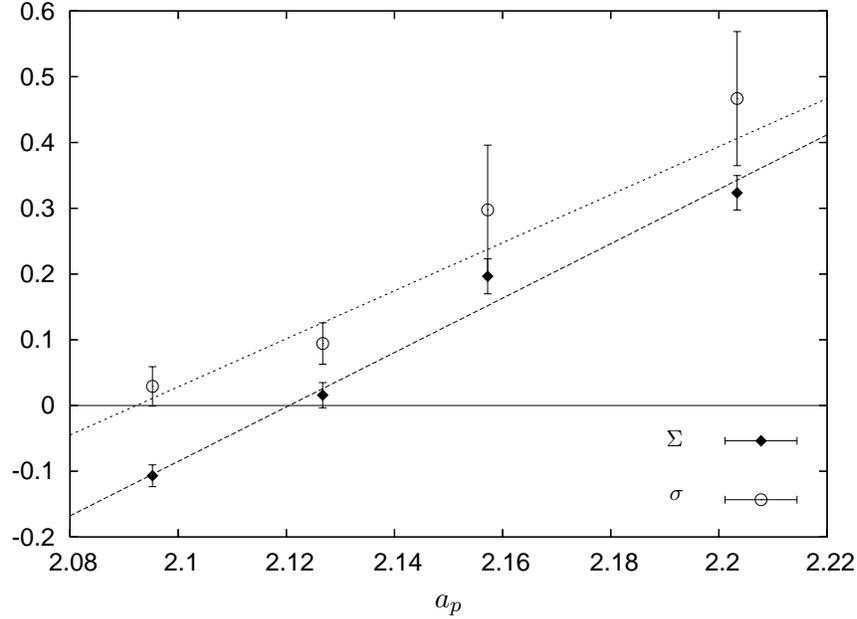}
\caption{Plot of the fluctuation surface tension $\sigma$ and of the mechanical surface tension $\Sigma$ as functions of the projected area per amphiphile $a_p$, for the $N=1152$ system.  The lines are linear fits to the two sets of data. The two sets of data appear to be clearly shifted and tilted  with respect to each other.}
\label{compare_sS}
\end{figure}

Comparison of tables \ref{tabS} and \ref{tabsig}, and inspection of figure \ref{compare_sS}, clearly indicate that the curves $\Sigma(a_p)$ and $\sigma(a_p)$ exhibit different slopes.
Furthermore, if one plots on the same figure $\sigma$ and $\Sigma$ as functions
of $a_p$, in the range where $\Sigma$ is linear with respect to $a_p$, the two data sets appear clearly shifted the one respect to the other, 
see figure \ref{compare_sS}, where these two quantities have been plotted 
for the $N=1152$ case.

\begin{figure}[h]
\center
\psfrag{ap}[ct][ct][1.]{$a_p$}
\psfrag{a}[bl][bl][1.][-90]{$a$}
\psfrag{x3}[br][br][.8]{$N=1152$}
\psfrag{xm}[br][br][.8]{$N=768$}
\psfrag{x2}[br][br][.8]{$N=512$}
\includegraphics[width=12cm]{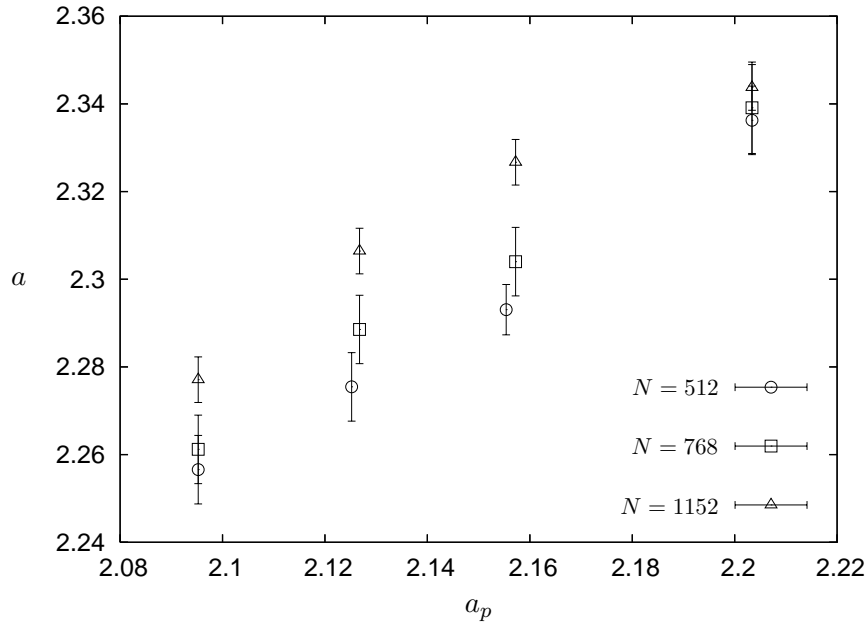}
\caption{Plot of the measured effective area per amphiphile $a$, as a function of the projected area per amphiphile $a_p$ (simulation box area), for the three values of $N$ here considered. }
\label{effa}
\end{figure}

In order to relate the two surface tension $\sigma$ and $\Sigma$, 
we need first to define the statistical ensemble which characterizes the system considered here, i.e., the minimal set of thermodynamical variables which fully describe the system state. The classical argument that the system area readjusts to its preferred value (see, e.g., \cite{Saf}) cannot be invoked here because of the constrain represented by the frame area (simulation box area).
The effective bilayer area  $A$ can be evaluated
using a discrete version of the formula
\begin{equation}
A=\int_{A_p} \de x\de y \sqrt{1+\p{\nabla h}^2},
\label{defA}
\end{equation}
where the field $h(x,y)$ is sampled during the MD simulations.
The effective area per amphiphile $a$ is plotted in fig.~\ref{effa} as a function of $a_p$. Inspection of
this figure clearly indicates that the effective area $a$ is not independent of the projected area $a_p$, but is rather a function of $a_p$.
Thus, the thermodynamical ensemble characterizing the framed bilayer here considered is the $(T,A_p=L^2, N)$ ensemble, where $L$ is the size of the simulation box face parallel to the bilayer.

The free energy in the $(T,A_p=L^2,N)$ ensemble for a bilayer membrane system is 
\begin{eqnarray}
F&=&-k_B T\ln Z=-k_B T\ln\p{\int {\mathcal D} h\,  e^{-\beta H(\{h\})} }\nonumber\\
 &\simeq& -k_B T\ln\p{ \prod_{i=0}^{N-1}\int \frac{ d h_i}{\lambda} e^{-\beta H(\{h_i\})} }\nonumber \\
       &=&-k_B T\ln \p{ \prod_{\bq}\int  \frac{d h_{\bq}}{\sqrt N a_p\lambda} e^{-\beta H(\{h_q\})} }\nonumber \\
    &=&\sigma A_p-k_B T\ln \pq{ \prod_{\bq}\int  \frac{d h_{\bq}}{\sqrt N a_p\lambda}
 e^{-  \frac \beta{2 L^2} \sum_{\bq}(\sigma q^2+ \kappa q^4)  |h_\bq|^2 } }\nonumber \\
    &=&\sigma A_p-k_B T \ln \pq{ \prod_{\bq} \frac{2\pi}{\lambda^2 a_p \beta (\sigma q^2+ \kappa q^4)} }^{\frac 1 2}\nonumber \\
    &=&\sigma A_p+\um k_B T\sum_\bq\ln \pq{\frac{\lambda^2 a_p (\sigma q^2+ \kappa q^4)}{2\pi k_B T} }\, .
\label{free}
\end{eqnarray}
The parameter $\lambda$ has the dimension of a length and is the analogous of the De Broglie wave-length in the ideal gas partition function. Without introducing
this parameter, the partition function would be dimensionful.
The equation of state for the intrinsic area is
\begin{eqnarray}
\media A&=&\dep {F(T,A_p)}{\sigma} =A_p+\frac 1 2 k_B T \sum_\bq \frac 1 {\sigma+\kappa q^2},
\label{area}
%\\ &=&A_p\pq{1+\frac{k_B T}{8\pi\kappa}\ln \p{\frac {\sigma+\kappa \qM^2}{\sigma+\kappa \qm^2}}}\, 
\end{eqnarray} 
which is a well known result, see, e.g, \cite{Rou}.
Let us now introduce the quantity  $N'$, which represents the number of membrane patches which fluctuate independently ($N'\le N/2$): the summations in equations (\ref{free}) and (\ref{area}) run over these $N'$ wavemodes, and thus we have $\sum_{\bq}=N'$.
Note that, in the large $A_p=L^2$ limit, $N'$ can be estimated as follows
\begin{eqnarray}
N'&=& \sum_{\bq}\simeq \p{\frac{L}{2 \pi}}^2 \int \de q^2 =\frac{L^2}{2 \pi} \int^\qs_\qm  q \de q\nonumber \\
&=& \frac{L^2}{4 \pi} \p{\qs^2-\qm^2} \simeq \frac{A_p}{4 \pi}, 
\label{evaln1}
\end{eqnarray} 
where we have taken $\qs\simeq 1$ (see discussion is section \ref{mis_s}), and $\qs\gg\qm\simeq 2 \pi/L\sim 0$, in the large system size limit.
Taking into account that $q^2=(2  \pi)^2 (n_x^2+n_y^2)/L^2$,  and since the surface tension $\Sigma$ is the thermodynamic conjugate of $A_p=L^2$, as defined by eq.~(\ref{defSigma}), we have
\begin{eqnarray}
\Sigma(T,A_p)&=&\D{F}{A_p}=\sigma -\um \frac{k_B T}{A_p} \sum_{\bq} \frac{\kappa q^4}{\kappa q^4+\sigma q^2}\\
      &=&\sigma +\um \frac{k_B T}{A_p} \sum_{\bq} \frac{\sigma }{\kappa q^2+\sigma } -\frac{k_B T}{2} \frac{N'}{A_p}\label{Sigma2}\\
%      &=& \sigma +\um \frac{k_B T}{(2\pi)^2} \sigma \int \frac{\de ^2 q }{\kappa q^2 +\sigma} -\frac{k_B T}{2} \frac{N'}{A_p}\label{Sigma3}\\
      &=& \frac{\media A}{A_p} \sigma -\frac{k_B T}{2} \frac{N'}{A_p}\label{Sigma4}\, ,
\end{eqnarray} 
where we substituted eq. (\ref{area}) into eq. (\ref{Sigma2}).
Equation (\ref{Sigma4}) relates thus the frame surface tension $\Sigma$ with 
two of the variables which define the actual ensemble $T,A_p$, and with the fluctuation surface tension $\sigma$.
Comparison of tables \ref{tabS} and \ref{tabsig} indicates that, for a given system size,  the 
slope of $\Sigma(a_p)$ is larger than  the slope of $\sigma(a_p)$: since 
$A>A_p$ by definition of intrinsic area, the first term on the rhs of eq.~(\ref{Sigma4}) accounts for the difference of the slope between the two curves.
Note that the second term on the rhs of eq.~(\ref{Sigma4}) takes into account 
the effect of the entropic elasticity on the system surface tension. This term
is proportional to the system temperature, and inversely proportional
to the system effective area per amphiphile $a_p'=A_p/N'$: the analogy with 
the case of polymers can be immediately drawn.
It is worth noting that a similar results was obtained by Farago and Pincus in ref.~\cite{FP}, where a continuous expression for the free energy (\ref{free}) was used, and where the surface tension $\sigma$ was taken to depend explicitly on the effective area $A$.
A first rapid check of equation~(\ref{Sigma4}) can be done by noting that it predicts that, if the surface tension $\sigma$ vanishes, the frame 
surface tension $\Sigma$ is negative. This is confirmed by inspection of figures \ref{compare_S} and \ref{compare_s}.

\begin{figure}[h]
\psfrag{ap}[br][br][1.]{$a_p$ }
\psfrag{Sigma}[br][br][.8]{$\Sigma$ }
\psfrag{sigma}[br][br][.8]{$\Sigma(T,A_p,\sigma)$}
\center
\includegraphics[width=12cm]{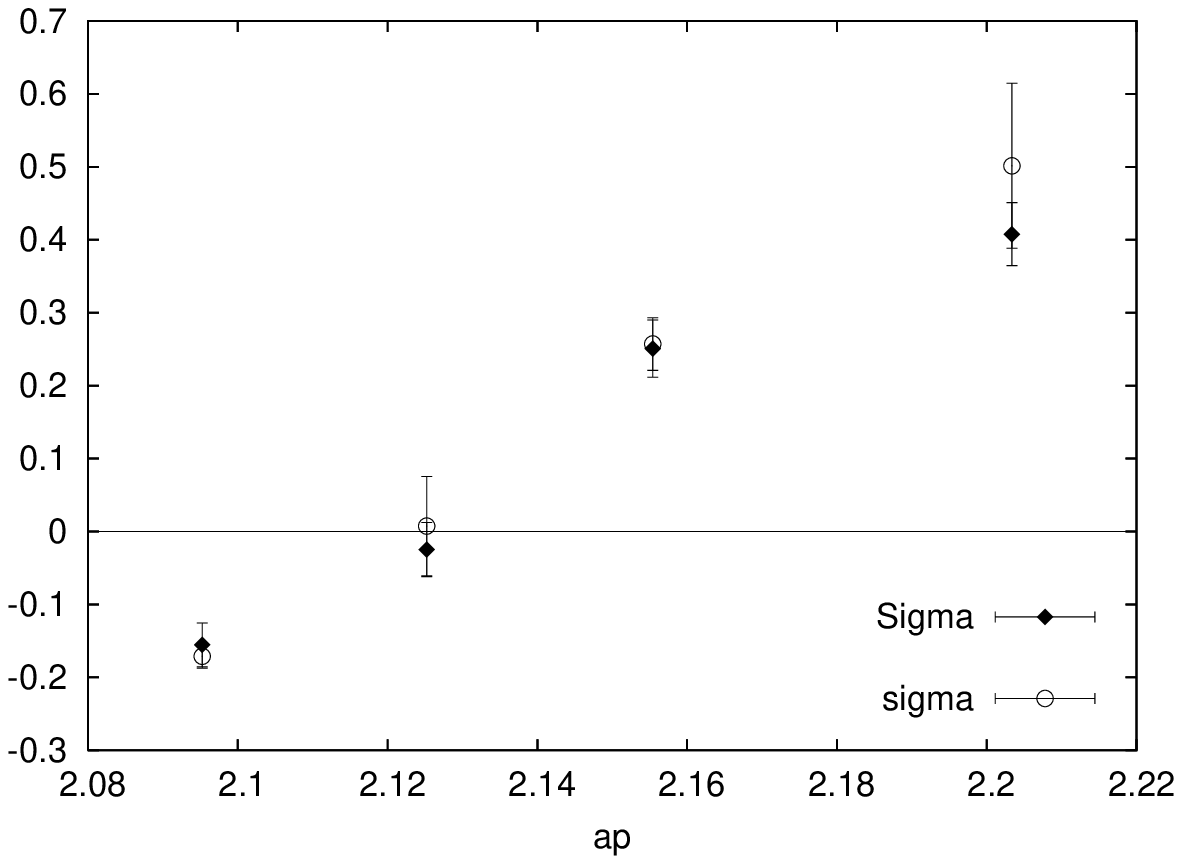}
\caption{Plot of the measured surface tension $\Sigma$ and of the estimate of the same quantity, as given by eq. (\ref{Sigma4}), as functions of $a_p$, for a system with $N=512$ amphiphiles.}
\label{simul1}
\end{figure}

\begin{figure}[h]
\psfrag{ap}[br][br][1.]{$a_p$ }
\psfrag{Sigma}[br][br][.8]{$\Sigma$ }
\psfrag{sigma}[br][br][.8]{$\Sigma(T,A_p,\sigma)$}
\center
\includegraphics[width=12cm]{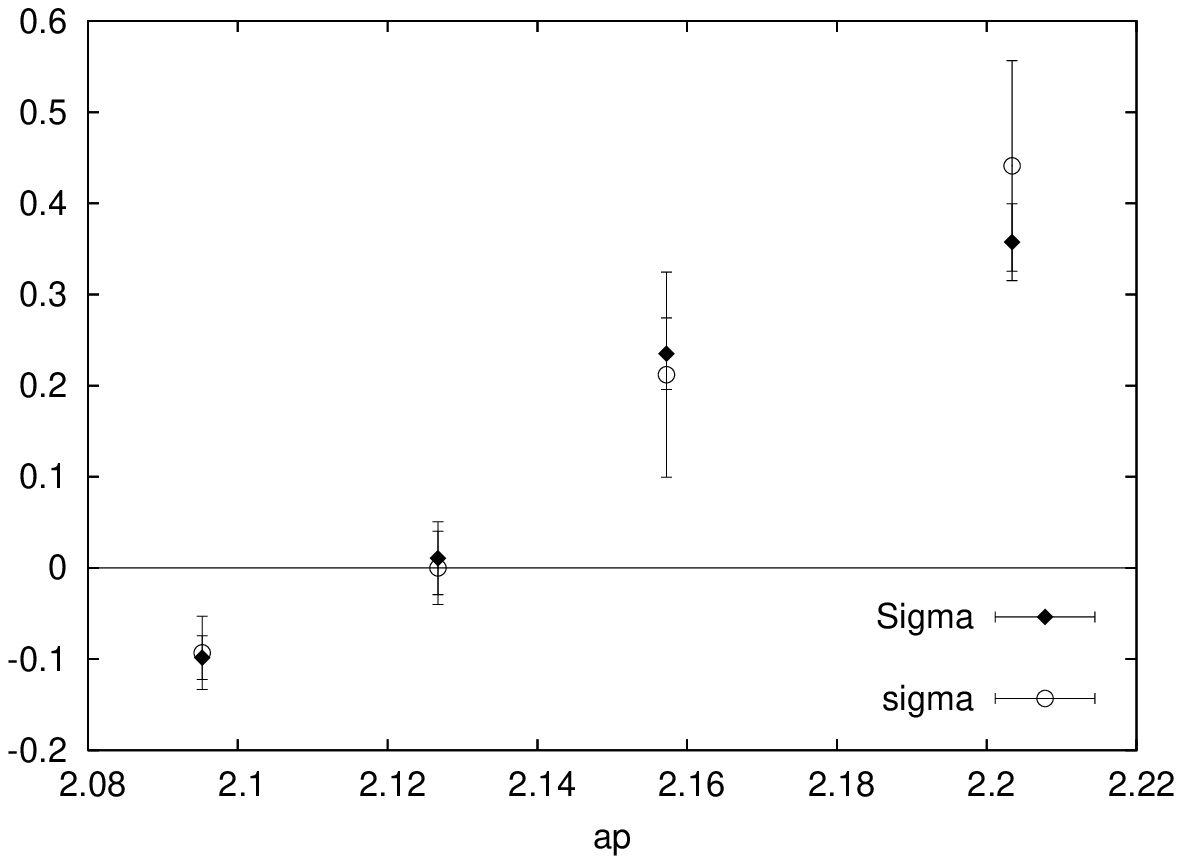}
\caption{Plot of the measured surface tension $\Sigma$ and of the estimate of the same quantity, as given by eq. (\ref{Sigma4}), as functions of $a_p$, for a system with $N=768$ amphiphiles.}
\label{simulm}
\end{figure}

\begin{figure}[h]
\psfrag{ap}[br][br][1.]{$a_p$ }
\psfrag{Sigma}[br][br][.8]{$\Sigma$ }
\psfrag{sigma}[br][br][.8]{$\Sigma(T,A_p,\sigma)$}
\center
\includegraphics[width=12cm]{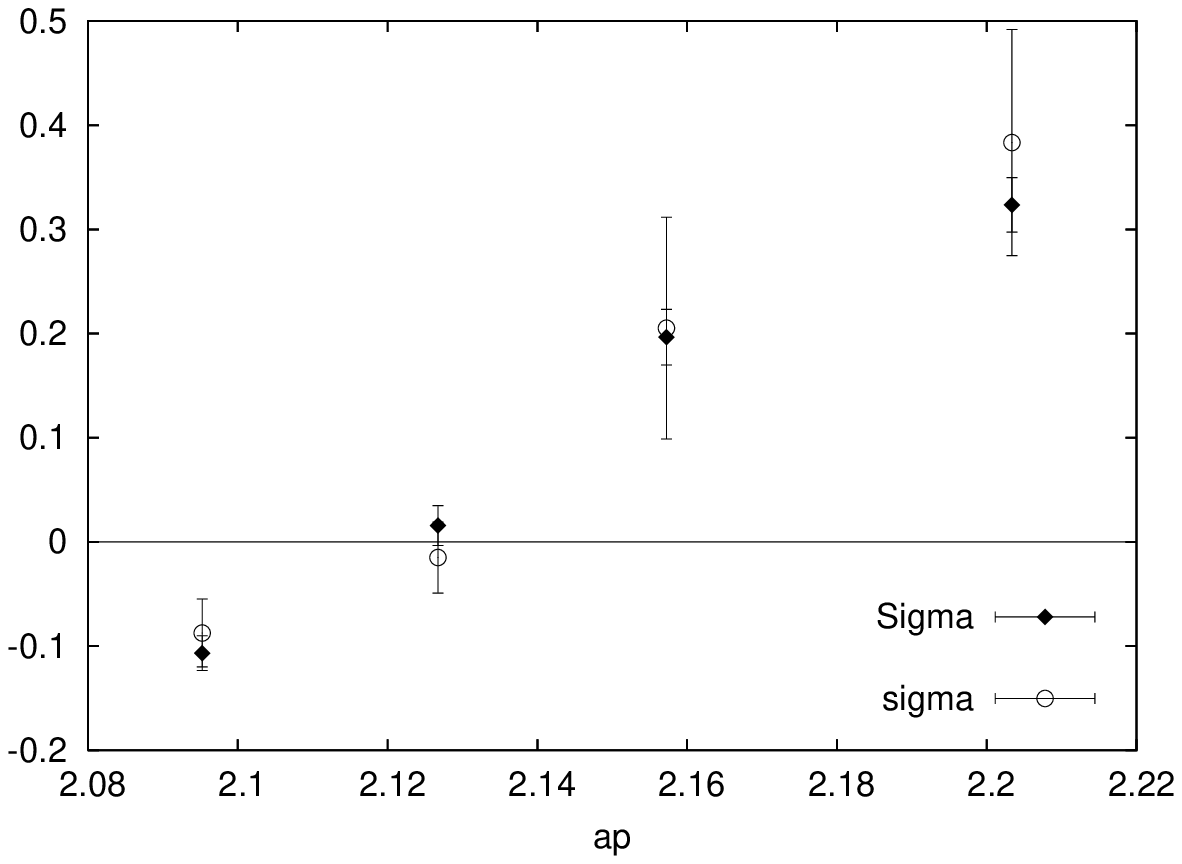}
\caption{Plot of the measured surface tension $\Sigma$ and of the estimate of the same quantity, as given by eq. (\ref{Sigma4}), as functions of $a_p$,  for a system with $N=1152$ amphiphiles.}
\label{simul2}
\end{figure}

In equation~(\ref{Sigma4}) the only adjustable parameter is $N'$, while $\sigma$ can be estimated as described in section  \ref{mis_s}, and $A$ can be 
sampled during the MD simulations using eq.~(\ref{defA}).
One can thus compare the values for $\Sigma$ predicted by eq.(\ref{Sigma4}) with
those directly measured as described in section \ref{mis_S}.
In the following, the optimal value of the parameter $N'$ will be determined by
fitting eq.~(\ref{Sigma4}) to the measured values plotted in fig.~\ref{compare_S}, for the three values of $N$ here considered.
With this fit we find $N'=155$ for $N=512$, $N'=181$ for $N=768$ and $N'=213$ for $N=1152$.
Using this values for $N'$, we plot the measured and the calculated values 
for $\Sigma$, as a function of $a_p$, for $N=512$ fig.~\ref{simul1}, for $N=768$ fig.~\ref{simulm}, and for
$N=1152$ fig.~\ref{simul2}. Inspection of these figures indicates a good agreement
between the values of $\Sigma$ predicted by eq.(\ref{Sigma4}) and those calculated as described in section \ref{mis_S}.

\begin{figure}[h]
\center
\psfrag{N}[ct][ct][1.]{$N$}
\psfrag{N1}[br][br][1.][-90]{$N'$}
\psfrag{Ap}[ct][ct][1.]{$A^*_p$}
\includegraphics[width=12cm]{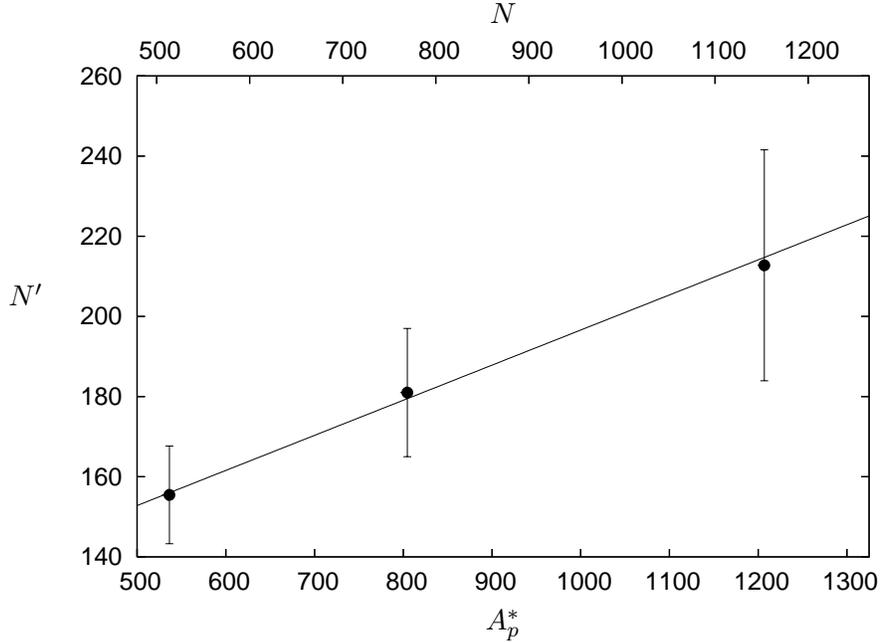}
\caption{Plot of the effective fluctuation modes $N'$ as a function of the vanishing tension projected area $A^*_p$ ($\sigma=0$) and  as a function of the number of molecules $N$ (upper $x$-axis). The line is a linear fit of the data, whose slope, as a function of $A^*_p$, is $0.086\pm 0.003\simeq 1/(4 \pi)$, as predicted by eq.~(\ref{evaln1}). }
\label{n1}
\end{figure}

We now consider the scaling behaviour of the quantity $N'$ as a function of the system size. In fig. \ref{n1}, $N'$ is plotted as a function of the tensionless  projected area $A^*_p$ (projected area at which $\sigma=0$), and as a function of the number of molecules $N$. Inspection of this figure, suggests that $N'$ is a linear function of $A^*_p$, and the slope of
such function is in good agreement with that predicted by eq.~(\ref{evaln1}). It is worth noting that, as discussed is section \ref{mis_s}, the projected area per molecule 
$a^*_p=A^*_p/(N/2)$ is independent of the system size, and thus we have $N'\sim A^*_p/(4 \pi)\sim N$, as one would expect in the large $N$ (or large $A_p$) limit.

Note that the intercept of the line plotted in fig.~\ref{n1} is non-zero, while eq.~(\ref{evaln1}) predicts a vanishing intercept in the large projected 
area limit.
However the argument used to obtain eq.~(\ref{evaln1}) is no longer valid 
for small value of $A_p$ ($N$), and thus in this limit such an equation is incorrect.
\section{Conclusion}\label{concl}
In the present paper we have measured both the surface tension $\sigma$ appearing in the Hamiltonian which governs the bilayer shape fluctuations (\ref{fapp}), and the surface tension $\Sigma$ which governs the elastic response of the system to
a change in its projected area $A_p$.
Using computer simulations, we have measured independently  these two quantities for bilayer with different projected area.  Our results indicate that the two surface tensions have different values for  a given value of  $A_p$. In particular
the two projected area per amphiphile $a^*_p$ and $a^\dagger_p$ are different, 
 where $a^*_p$ is the  projected area per amphiphile at which the surface tension $\sigma$ vanishes, while $a^\dagger_p$ is the projected area per amphiphile where the surface tension $\Sigma$ vanishes.
Furthermore, the two surface tensions are found to exhibit different slopes as 
functions of the projected area per molecule $a_p$, for a given system size.
Using a simple thermodynamic argument, we manage to relate the two quantities 
$\sigma$ and $\Sigma$, and the relation between them which we found, eq.~(\ref{Sigma4}),
nicely fits the data that we obtain from simulations.  This is the most important result of the present work: it indicates that eq.~(\ref{Sigma4}) succeeds to capture the basic relation between the two quantities. 
%It is important to remark that, to the best of our knowledge, in no other work the two surface tensions associated to the two different types of deformation, which a membrane can undergo, have been contemporaneously measured and compared.
Note that in a previous work \cite{Ste}, where computer simulations of model membrane were considered, the two surface tensions were found to be proportional the one respect to the other, but no argument was introduced to explain such an effect.

The difference between the two quantities, does not appear to be due to some 
size effect, since the vanishing areas  $a^*_p$ and $a^\dagger_p$ 
do not depend on the system size. 
The scaling behaviour of the effective wavemode number $N'$  as a function 
of the system size is a rather delicate issue. In the present work we find that $N'$ scales linearly with the number of molecule $N$ (the tensionless area $A^*_p$), but in a restricted range of values of $N$ ($A_p$). We were limited in the choice of the values of $N$ by the considerable amount of computation time required by the simulations. 
For larger values of $N$ one would expect that $N'$ grows either as $N$ or more slowly than $N$. If this were the case, since $A_p\sim N$, the second term of the rhs of
eq.~(\ref{Sigma4}) would vanish in the large $N$ limit. In this case eq.~(\ref{Sigma4}) would read $\Sigma\simeq A/A_p \cdot \sigma$. We plan to consider larger system size in a future work.
%The ratio $N'/A_p$, appearing on the rhs of 
% eq.~(\ref{Sigma4}), actually decreases as the number of amphiphiles $N$ increases, an accurate scale analysys will be addressed elsewhere.
%

In conclusion, the results contained in this paper puts in evidence that in a bilayer membrane, the
surface tension $\sigma$  
 appearing in the elasticity Hamiltonian  (\ref{fapp}) and the mechanical
surface tension $\Sigma$ are two distinct thermodynamic quantities, which have to be measured independently for a full characterization of the system elastic properties, although for a constrained system, the geometrical constraints impose a relation between them.

It would be interesting to measure these two quantities in a real system, e.g, a lipid vesicle manipulated with a micropipette. On the basis of the results presented in this paper, one would expect that the geometrical constrains imposed by the micropipette would lead to a difference between the two surface tensions.
\begin{acknowledgments}
The author is indebted to  R. Lipowsky for introducing him to this  topic, for many interesting discussions and for a critical reading of this manuscript.
I am grateful to J.B. Fournier for the long and interesting discussions on the two tension issue.
I also thank L. Peliti for many discussions, and for his  encouragements.
This work was partially supported by MIUR-PRIN 2004.
Finally, I thank the CRdC AMRA for the use of its computational resources.
\end{acknowledgments}

\newpage 

\end{document}